# Correlating the Electronic Structures of Metallic/Semiconductor MoTe$_2$ Interface to its Atomic Structures

Bo Han[1,2], Chen Yang[3,4], Xiaolong Xu[3], Yuehui Li[1], Ruochen Shi[1], Kaihui Liu[3], Haicheng Wang[5*], Yu Ye[3,6*], Jing Lu[3,4,6], Dapeng Yu[3,7], and Peng Gao[1,6*]

[1] *Electron Microscopy Laboratory and International Center for Quantum Materials, School of Physics, Peking University, Beijing 100871, China*

[2] *Department of Material Physics and Chemistry, University of Science and Technology Beijing, Beijing 100083, China*

[3] *State Key Laboratory for Artificial Microstructure & Mesoscopic Physics, School of Physics, Peking University, Beijing 100871, China*

[4] *Academy for Advanced Interdisciplinary Studies, Peking University, Beijing 100871, China*

[5] *State Key Laboratory of Advanced Materials for Smart Sensing, GRINM Group Co. Ltd., Beijing 100088, and GRIMAT Engineering Institute Co. Ltd., Beijing 101402, China*

[6] *Collaborative Innovation Center of Quantum Matter, Beijing 100871, China*

[7] *Shenzhen Key Laboratory of Quantum Science and Engineering, Shenzhen 518055, China*

*Corresponding authors: hcwang@grinm.com, ye_yu@pku.edu.cn, and p-gao@pku.edu.cn

# Abstract

Contact interface properties are important in determining the performances of devices based on atomically thin two-dimensional (2D) materials, especially those with short channels. Understanding the contact interface is therefore quite important to design better devices. Herein, we use scanning transmission electron microscopy, electron energy loss spectroscopy, and first-principles calculations to reveal the electronic structures within the metallic (1T')-semiconducting (2H) $MoTe_2$ coplanar phase boundary across a wide spectral range and correlate its properties and atomic structure. We find that the 2H-$MoTe_2$ excitonic peaks cross the phase boundary into the 1T' phase within a range of approximately 150 nm. The 1T'-$MoTe_2$ crystal field can penetrate the boundary and extend into the 2H phase by approximately two unit cells. The plasmonic oscillations exhibit strong angle dependence, i.e., a red-shift of π+σ (approximately 0.3 eV–1.2 eV) occurs within 4 nm at 1T'/2H-$MoTe_2$ boundaries with large tilt angles, but there is no shift at zero-tilted boundaries. These atomic-scale measurements reveal the structure-property relationships of 1T'/2H-$MoTe_2$ boundary, providing useful information for phase boundary engineering and device development based on 2D materials.

Two-dimensional (2D) transition metal dichalcogenides (TMDs) have attracted extensive attention due to their potential applications in nanoelectronics [1,2]. In these atomically thin TMDs devices, contact interface properties can significantly influence the performance, especially in short-channel devices [3,4]. An imperfect interface between the electrode and a 2D semiconducting TMD can cause Fermi level pinning and thus result in high resistance across the contact [5,6], which limits potential applications as device sizes scale down. Recent strategies such as indium/gold contacts [7], tunneling contacts [8], and metallic 2D material contacts [9] have been used to reduce contact resistance in long-channel devices [3,7–10]. However, these techniques are less effective in short-channel devices or large-scale applications. Recently, heterophase (e.g. metallic 1T'-$MoTe_2$ [11,12] and semiconducting 2H-$MoTe_2$ [13,14]) coplanar [15–17]) structures have been demonstrated to effectively reduce contact resistances in stable integrated circuits [18] by avoiding introduction of defects and impurities from step-by-step device fabrication processes [19–22]. These keep the promise of phase engineering as an effective way to reduce short-channel device contact resistances in order to achieve the low contact resistance requirements of the International Technology Roadmap for Semiconductors [4].

The properties of these coplanar boundaries (e.g., 1T'/2H-$MoTe_2$) should be dictated to their atomic structures, such as the interfacial sharpness, relative orientation between metallic and semiconducting phases, and nature of the interfacial bonds, which, unfortunately, remain largely unknown due to a lack of techniques that correlate the electronic structures of atomically thin interfaces to their microstructures. Conventional optical measurements generally offer neither sufficient spatial resolution to probe the local properties of interfaces and defects nor the ability to determine their atomic structures. Scanning tunneling microscopy/spectroscopy (STM/STS) techniques are typically sensitive only to the energy density of states (DOS) near the Fermi level with respect to TMDs interfaces [23–26]. As a result, the dependence of deep ultraviolet (DUV) range plasmonic properties and inner shell transitions on atomic structure has rarely been investigated.

Here, we use advanced monochromatic energy loss spectroscopy (EELS) with a scanning transmission electron microscopy microscope (STEM) with sub-10 meV energy and atomic spatial resolutions to study 1T'/2H-MoTe$_2$ phase boundaries. We correlate the atomic structure of each phase boundary with its electronic states over a wide spectral range from hundreds of meV to hundreds of eV. We find that the interband transition behavior of MoTe$_2$ exhibits delocalized character within approximately 150 nm at all 1T'/2H phase boundaries for various tilt angles (relative orientations). The DUV plasmon oscillation ($\pi+\sigma$) peak has a red-shift of approximately 0.3 eV–1.2 eV within 4 nm of the boundary at large tilt angles due to a change in the dielectric function and decreased free electron density. No substantial shift is observed for those boundaries with small tilt angle, which indicates that the relative orientations of the two crystal grains have significant influence on the contact properties. Furthermore, the interactions between 1T' and 2H phases change the crystal fields at all phase boundaries and thus alter the energy-loss near-edge structures (ELNES) of the Te-N and Te-M edges within approximately two-unit cells of the boundary on the 2H-MoTe$_2$ side. These findings of microstructure-dependent electronic structures at 1T'/2H-MoTe$_2$ phase boundaries can enable us to understand device contact properties and further guide researchers in designing high-performance nanodevices via coplanar boundary engineering.

Figs. 1(a)–1(b) show the atomic structures of the 2H-MoTe$_2$ and 1T'-MoTe$_2$ phases. Unlike 2H-MoTe$_2$, in which the Mo and Te atoms have a regular prismatic arrangement, 1T'-MoTe$_2$ exhibits a distorted atomistic arrangement, i.e., one Te-atom layer is offset from the next, resulting in octahedral coordination structures arranged

around Mo atoms [Fig. 1(c)]. This can form multiple types of phase boundaries, as shown in Figs. 1(d)–1(e). Each 1T'/2H phase boundary has two angle parameters: the angle $\varphi$ between the zigzag ($<11\bar{2}0>$) direction of the 2H phase and the boundary plane or line and the tilt angle $\theta$ between the zigzag of the 2H phase and the [010] of the 1T' phase. We label only the latter parameter $\theta$ since the acquired data shows no clear dependence on $\varphi$. In order to improve measurement accuracy, tilt angles are determined in reciprocal space, as shown in Figs. 1(f) and 1(g). The measurement uncertainties are discussed in detail in Fig. S1 in the Supplemental Material.

The intrinsic valence electron energy-loss spectra (VEELS) shown in Fig. 2(a) demonstrate various valence electron transitions between 2H-MoTe$_2$ and 1T'-MoTe$_2$. The VEELS of 2H-MoTe$_2$ contain five exciton peaks that represent its interband transitions [27] (see details in Fig. S2 in the Supplemental Material). Unlike 2H-MoTe$_2$, the spectrum of 1T'-MoTe$_2$ incorporates only one broad peak. This is caused by the absence of an energy gap near the Fermi level, as per the calculated DOS shown in Fig. 2(b). The bandgap of semiconducting 2H-MoTe$_2$ is 0.9 eV (see Fig. S2 in the Supplemental Material), which is consistent with previous optical measurements [11,27,28]. Fig. 2(c) shows that a series of evanescent peaks extends across the phase boundary from 2H-MoTe$_2$ to 1T'-MoTe$_2$ (~60° or 0° in tilt). This behavior might stem from that the interband transition of valence electrons creates electron-hole pairs of which the electric field is a long-range Coulomb interaction, as well as the delocalization effect [29]. Fig. 2(g) shows that the interaction range is fitted to be approximately 150 nm around the phase boundary (the fitting method is in the Supplemental Material). Measurements from other phase boundaries with different tilt angles show that the typical interaction width is approximately 100 nm–150 nm and no distinguished angle dependence [Figs. 2(h)–2(k) and Fig. S3 in the Supplemental Material].

STEM-EELS has the ability to probe plasmon oscillations in the DUV range (> approximately 5 eV) with ultra-high spatial resolution. Figs. 3(a) and 3(b) show plasmon modes collected from a 6 nm×16 nm area that contains a ~0° MoTe$_2$ phase boundary. Two dominant peaks π and π+σ can be observed in the 5 eV–35 eV energy loss range [30], which is consistent with the theoretical calculations in Fig. S4 in the Supplemental Material. In contrast to the small energy shift of the ~0° phase boundary, the energy shift and intensity change at the ~8.6° phase boundary are substantial. As shown in Figs. 3(c)–3(e), the DUV plasmon oscillation π+σ peak has a red-shift of approximately 0.75 eV within 4 nm of an ~8.6° phase boundary. The peak intensity also decreases within the ~8 nm region. This is likely due to defective bonds at the phase boundary.

The energy loss of the π+σ plasmon oscillation peak $E_p$ is determined using

$$E_p = \hbar \sqrt{\frac{ne^2}{\varepsilon_0 m}}$$

where $n$ represents the density of free charges, $e$ is the electron charge, $\varepsilon_0$ is the permittivity of free space, and $m$ represents the effective electron mass. The red-shift of the π+σ plasmon mode at the phase boundary is attributed to a reduction in the effective electron density at the phase boundary. This is consistent with the Kramers-Kronig (K-K) analysis in Figs. 3(f) and 3(g), as well as Fig. S5 in the Supplemental Material. In order to identify the angle-dependent electronic properties of MoTe$_2$ phase boundaries, we investigated various tilt angles [Fig. 3(h)]. At boundaries with large tilt angles, the significant red-shift of the π+σ plasmon mode (19 eV) indicates weaker σ bonds (i.e., weak interactions between the σ electron clouds of Mo and Te atoms), which harm carrier injection. The subtle energy shift of the π+σ mode at the near zero-tilted phase boundary avoids this high carrier injection barrier. In this sense, the phase boundary tilt angle can be used as a knob to tune coplanar structure contact properties.

The inner shell electronic structure of the 1T'/2H-MoTe$_2$ phase boundary is also studied in Figs. 4(a)–4(c). The Te-N edge of 2H-MoTe$_2$ contains two peaks at 41 eV and 42.5 eV that are not well separated in the 1T' phase. Similarly, the Mo-N edge is more pronounced in 2H-MoTe$_2$ than in 1T'-MoTe$_2$. At various positions on both sides of the phase boundary, the Te-N edge of 2H-MoTe$_2$ is altered only two-unit cells away from the phase boundary. The corresponding fitting of 2H and 1T' components shown in Fig. 4(d) also confirms that the Te-N ELNES of 2H-MoTe$_2$ deviates from the intrinsic shape near the phase boundary. This subtle change in the Te-N edge may indicate that the crystal field of 1T'-MoTe$_2$ extends across the boundary into the 2H phase for two-unit cells. An analogous phenomenon can be observed on the Te-M edge. The fine peaks on the Te-M edge ELNES from 570 eV–630 eV stem from crystal field splitting of the Te-3d orbital. In the 1T' phase near the boundary, these peaks still remain but are broadened [Fig. 4(e)]. The Te-M ELNES of 2H-MoTe$_2$ near the boundary also deviates from its intrinsic shape, as shown in Fig. 4(f). The atomic structure of the phase boundary deviates from the normal perfect lattice, resulting in reconstruction of the crystal field in a localized region. Other phase boundaries with different tilt angles exhibit similar behaviors in the Te-N and Te-M ELNES and similar two-unit cell interaction ranges, as shown in Figs. S6 and S7 in the Supplemental Material.

Broken translational symmetry at structural defects is often accompanied by changes in electronic structures. Previous studies reported that the boundaries in TMDs materials could influence their optical and electronic properties [31–34], due to differences in the atomic arrangements between the boundaries and the bulk parent phase. In this work, we correlate the electronic structure with the atomic arrangements (tilt angle) and find that such microstructure (angle) dependent behaviors are different for different physical excitation processes, i.e., the angle dependence is insensitive to the excitonic and inner shell excitations but sensitive to the plasma oscillations. The interfaces between 1T'-MoTe$_2$ and 2H-MoTe$_2$ with large tilt angles show substantial red-shift, which is expected to introduce high carrier injection barriers. This may be due to imperfect interfacial atomic arrangements. Such a strong plasma oscillation angle dependence indicates that adjusting the relative orientations of the two crystal grains of a heterophase structure provides a new strategy for controlling boundary electronic structures and further tuning contact properties. Therefore, angle-controllable synthesis technologies in future may be used to make metal-semiconductor 2D heterostructures satisfy contact requirements in nanoelectronics.

In summary, we used monochromatic STEM-EELS with high spatial resolution and high energy resolution to study the atomic and electronic structures of 1T'/2H-MoTe$_2$ phase boundaries with various tilt angles across a wide spectral range. The VEELS of 2H-MoTe$_2$ incorporated five exciton peaks that extend through the boundary by 100–150 nm. The Te-N and Te-M core losses exhibited 1T'-MoTe$_2$ features for a distance of two-unit cells in the 2H phase, indicating that the 1T'-MoTe$_2$ crystal fields penetrated the boundary and extended a short distance into the 2H phase. Interestingly, the π+σ mode of DUV plasmon oscillations exhibited strong angle dependence. There is a red-shift of approximately 0.3 eV–1.2 eV within a 4 nm area for large titled phase boundaries, indicating change of dielectric function as well as the barrier for carrier injection. In contrast, no substantial shift is observed for near-zero and 60° tilted boundaries. Our atomic scale measurements using STEM-EELS help to elucidate the properties of coplanar metal-semiconductor contacts in TMDs and shed light on electrical and photoelectrical device design via phase boundary engineering.

This work was supported by the National Key R&D Program of China (2016YFA0300903 and 2018YFA0306900), the National Natural Science Foundation of China [11974023, 51672007, and 51971028], the Key R&D Program of Guangdong Province (2018B030327001, 2018B010109009, 2019B010931001), Bureau of Industry and Information Technology of Shenzhen (No. 201901161512), the National Equipment Program of China (ZDYZ2015-1) and the "2011 Program" from the Peking-Tsinghua-IOP Collaborative Innovation Center of Quantum Matter. We gratefully acknowledge the Electron Microscopy Laboratory at Peking University for the use of electron microscopes.

# Figures and captions

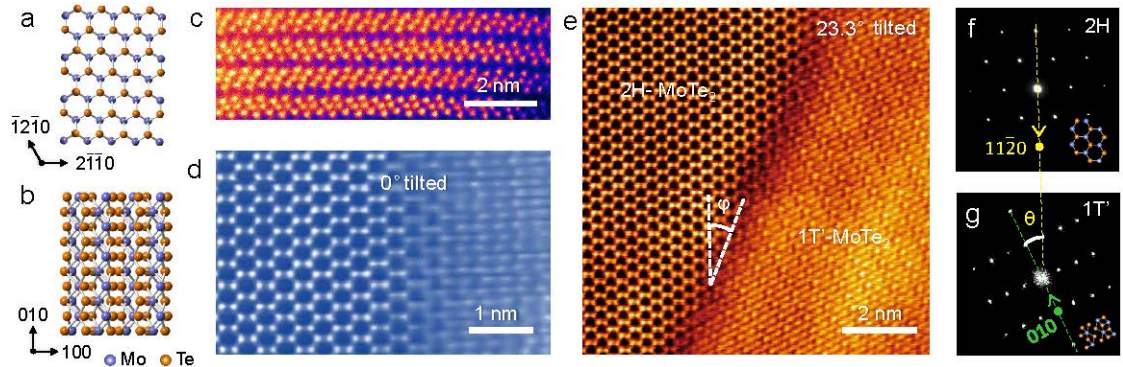

FIG. 1. Atomic structures of 1T'/2H-MoTe$_2$ coplanar heterostructures. (a) Atomistic models of (a) 2H-MoTe$_2$ and (b) 1T'-MoTe$_2$ viewed from [001]. (c) An atomically resolved STEM image of 1T'-MoTe$_2$ interlayers, viewed from the [010] zone axis. Atomically resolved STEM image of a MoTe$_2$ metallic (1T')/semiconductor (2H) coplanar heterojunction with tilt angles of (d) ~0° and (e) ~23.3°. (f) An electron diffraction pattern of 2H-MoTe$_2$ and (g) a fast Fourier Transform pattern of 1T'-MoTe$_2$. The tilt angle θ of the MoTe$_2$ heterostructure is determined by the zigzag direction of the 2H phase (yellow dashed line) and the [010] direction of the 1T' phase (green dashed line).

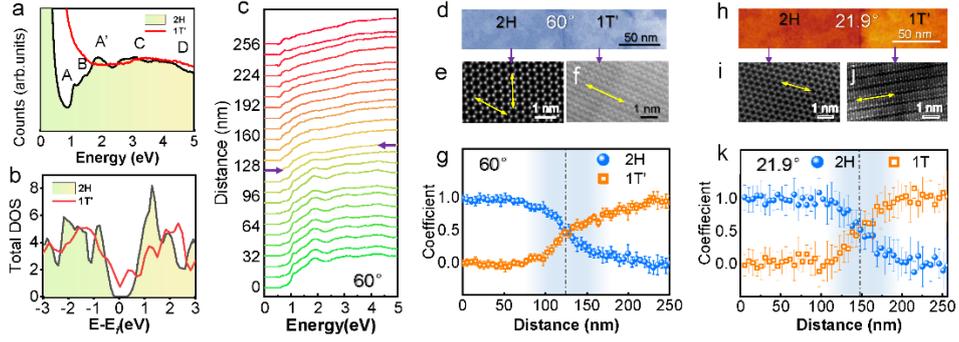

FIG. 2. VEELS of 1T'/2H-MoTe$_2$ phase boundaries. (a) VEELS of the intrinsic 2H (black line) and 1T' MoTe$_2$ (red line). (b) Calculated density of states (DOS) of 2H-MoTe$_2$ (black line) and 1T'-MoTe$_2$ (red line). Unit of vertical coordinates: states/ (eV. atom). 2H-MoTe$_2$ shows an intrinsic bandgap. (c) The VEELS series across the 1T'/2H phase boundary. The purple arrow indicates the location of the 1T'/2H-MoTe$_2$ phase boundary. (d) A HAADF image of a ~60° boundary showing the spectra-collected region (51 nm×256 nm), within which the title angle is determined using atomically resolved HAADF images of (e) 2H-MoTe$_2$ and (f) 1T'-MoTe$_2$. (g) The VEELS fitting coefficient at a 1T'/2H-MoTe$_2$ boundary versus distance. (h) A low-magnification HAADF image of a ~21.9° boundary and corresponding atomically resolved HAADF images of (i) 2H-MoTe$_2$ and (j) 1T'-MoTe$_2$; (k) the VEELS fitting coefficient at a ~21.9° 1T'/2H-MoTe$_2$ boundary. Blue sphere: 2H; Orange square: 1T'. The dashed gray line shows the location of the 1T'/2H MoTe$_2$ phase boundary.

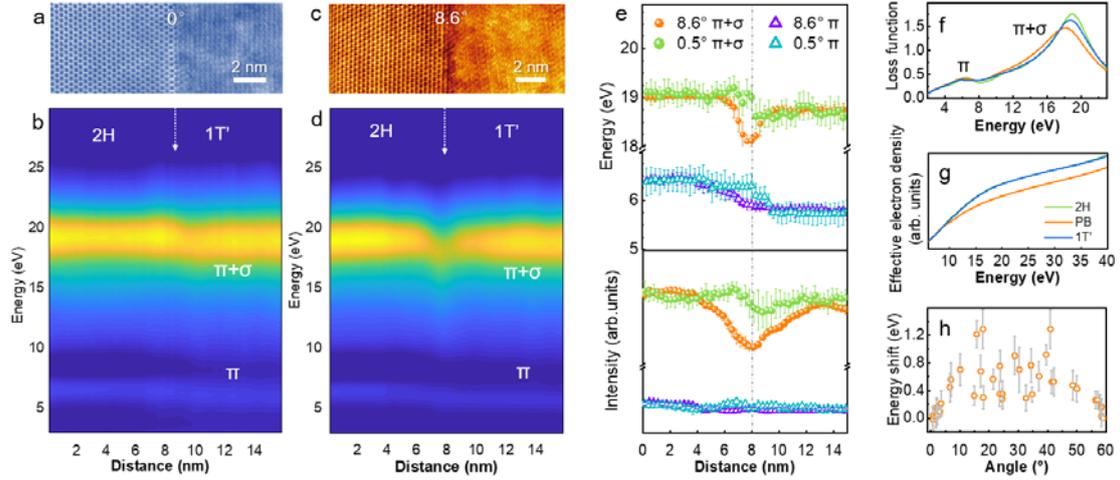

FIG. 3. Plasmon oscillations at 1T'/2H-MoTe$_2$ phase boundaries. (a) A HAADF image showing the spectra-collected region (6 nm×16 nm) containing a boundary with a tilt angle of ~0°. (b) The spatially resolved plasmon oscillation of the 1T'/2H-MoTe$_2$ boundary. The white, dashed arrow indicates the phase boundary location. The π-mode energy loss peaks from MoTe$_2$ are located at 6.4 eV (2H) and 5.7 eV (1T'). The energy value of the π+σ mode peak is ~19 eV. (c) A HAADF image and (d) the corresponding plasmon oscillation of the heterostructure with a tilt angle of ~8.6°. (e) Plasmon resonance peak energy values and intensities are plotted versus the distance across the phase boundary. The phase boundary is indicated by the gray dashed line and determined from the corresponding HAADF image. (f) The loss functions and (g) effective electron densities of 2H-MoTe$_2$ (green), 1T'-MoTe$_2$ (blue), and the phase boundary (orange). (h) The π+σ mode energy shifts at phase boundaries with various tilt angles. The error bars indicate standard deviations calculated from positions within 0.8 nm of the phase boundaries.

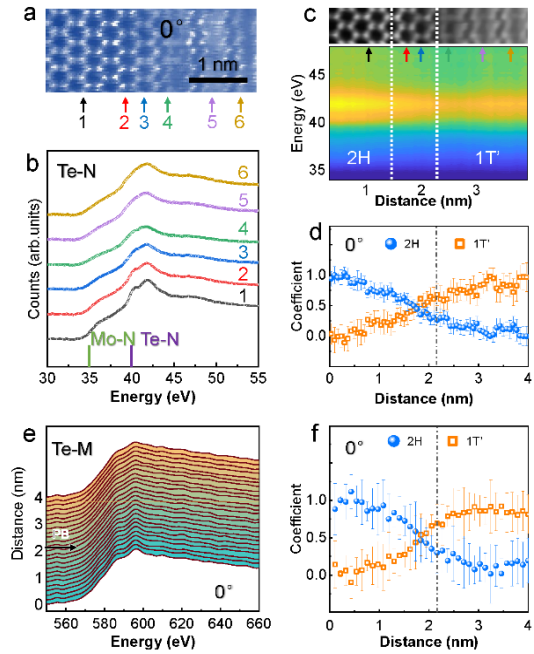

FIG. 4. ELNES at the 1T'/2H-MoTe$_2$ boundary. (a) A HAADF image showing the spectra-collected region. (b) Te-N ELNES at six locations are indicated by the black, red, blue, green, purple and yellow arrows. (c) A Te-N ELNES intensity map at the 1T'/2H boundary. The Te-N edge of 2H-MoTe$_2$ changes rapidly within approximately two-unit cells of the boundary (circled by white dashed lines). (d) The 1T' and 2H component fitting coefficients at the boundary. (e) The EEL spectra of Te-M ELNES at the 1T'/2H boundary and (f) corresponding fitting coefficients. The black arrow and black dashed lines indicate the location of the 1T'/2H-MoTe$_2$ phase boundary.

**Supplemental Material for：**

**Correlating the Electronic Structures of Metallic/Semiconductor MoTe$_2$ Interface to its Atomic Structures**

Bo Han[1,2], Chen Yang[3,4], Xiaolong Xu[3], Yuehui Li[1], Ruochen Shi[1], Kaihui Liu[3], Haicheng Wang[5*], Yu Ye[3,6*], Jing Lu[3,4,6], Dapeng Yu[3,7], and Peng Gao[1,6*]

[1] *Electron Microscopy Laboratory and International Center for Quantum Materials, School of Physics, Peking University, Beijing 100871, China*

[2] *Department of Material Physics and Chemistry, University of Science and Technology Beijing, Beijing 100083, China*

[3] *State Key Laboratory for Artificial Microstructure & Mesoscopic Physics, School of Physics, Peking University, Beijing 100871, China*

[4] *Academy for Advanced Interdisciplinary Studies, Peking University, Beijing 100871, China*

[5] *State Key Laboratory of Advanced Materials for Smart Sensing, GRINM Group Co. Ltd., Beijing 100088, and GRIMAT Engineering Institute Co. Ltd., Beijing 101402, China*

[6] *Collaborative Innovation Center of Quantum Matter, Beijing 100871, China*

[7] *Shenzhen Key Laboratory of Quantum Science and Engineering, Shenzhen 518055, China*

*Corresponding authors: hcwang@grinm.com, ye_yu@pku.edu.cn, and p-gao@pku.edu.cn

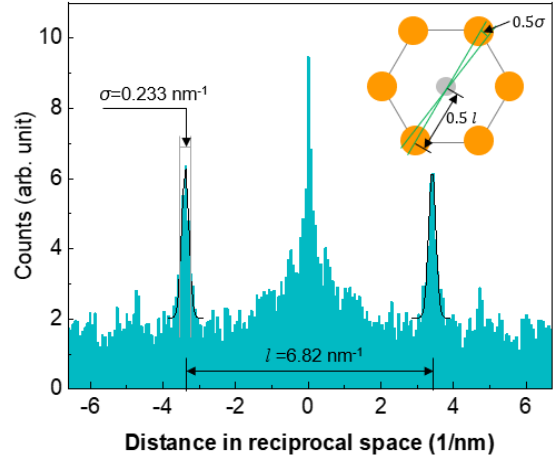

**FIG. S1. Uncertainties of tilt angles measurements.** The black plots represent the Gaussian function fitting results. Orange circles are the first order diffraction FFT pattern of a 2H-MoTe$_2$. Green lines indicate the line profiles for angle measurements.

In order to improve the measurement accuracy, the tilt angle is determined by FFT pattern of STEM images. The uncertainty mainly origin from the artificial measurement uncertainty in reciprocal space, i.e., the position difference of profiling lines between two diffraction spots.

We fit each diffraction spot with Gaussian function and its full width at half maximum (FWHM) is used to present the uncertainty.

The uncertainty of title angles can be calculated by

$$\triangle \theta \approx \sin\theta = \frac{\sigma}{l \pm \sigma}$$

where $\sigma$ is FWHM of Gaussian function for fitting a diffraction spot, $l$ represents the distance between two one-order FFT spots of 2H or 1T' MoTe$_2$

For instance, for a 2H-MoTe$_2$, $\sigma$ is measured to be 0.233 nm$^{-1}$, $l$ measured to be 6.82 nm$^{-1}$, as shown in Fig. R5. The uncertainty of tilt angle can be calculated by:

$$\triangle \theta \approx \frac{\sigma}{l \pm \sigma} = \frac{0.233}{6.82 \pm 0.233} \times 100\% = 3.42\% \pm 0.12\%$$

Therefore, this method of measuring tilt angles in reciprocal space should be relatively accurate.

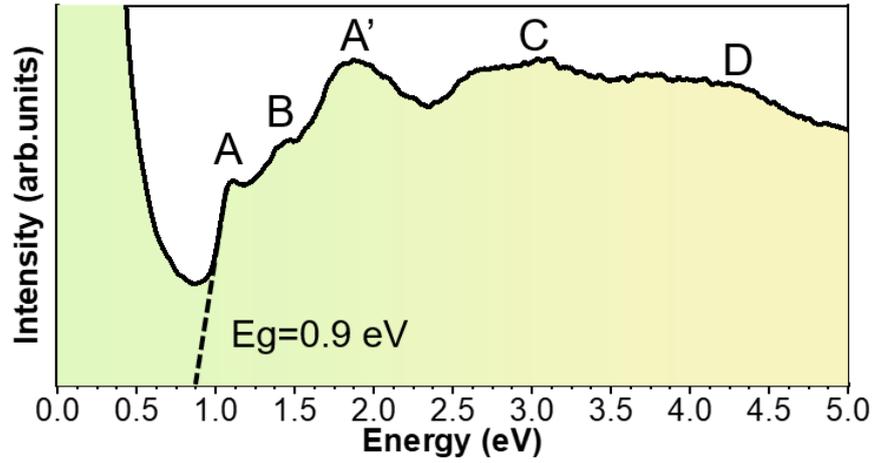

**FIG. S2. The band gap measurement of 2H-MoTe$_2$.**

The band gap of 2H-MoTe$_2$ is 0.9 eV on basis of VEELS. The characteristic exciton (electron-hole pairs) peaks A and B represent the direct transition at K point in the Brillouin zone. The energy difference between peak A and B is of 300 meV due to spin orbit coupling, which is consistent with optical characterization. Peak A' is caused by interlayer transition while peak C arises from the parallel bands at Γ point. [1]

It is worth mentioning that compare to the optical technique that only detects direct transition behavior, STEM-EELS has the advantages to mapping excitons at subwavelength scales [2] and probe the indirect transition of electrons and transition from deeper energy levels, which probably leads to the peak D at higher energy range (~4.3 eV).

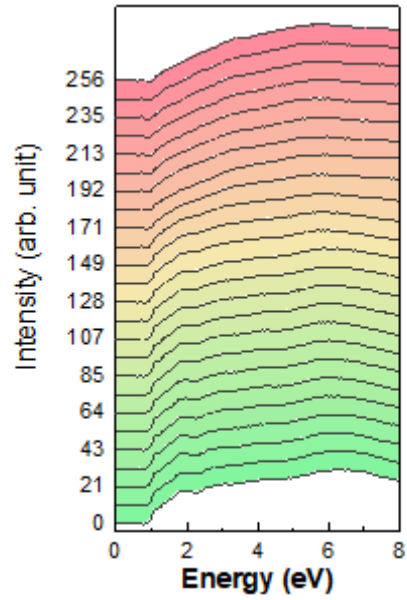

**FIG. S3.** Valence EELS series across ~21.9° tilted 1T'/2H-MoTe$_2$ phase boundary.

The VEELS of ~0° tilted 1T'/2H-MoTe$_2$ phase boundary indicates that the interaction range is ~100-150 nm across the phase boundary, which also shows the same behavior as the boundaries with large tilt angle.

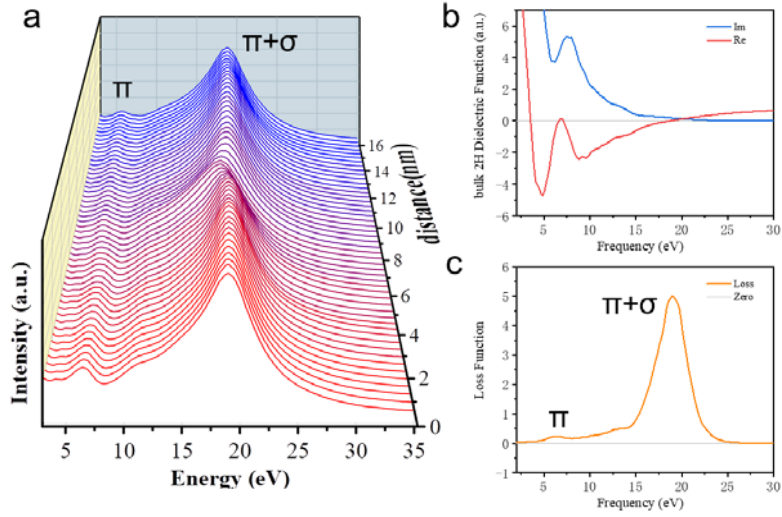

**FIG. S4. The calculated dielectric function of 2H-MoTe₂.** (a) Spatial resolved plasmon oscillation of 1T'/2H-MoTe2 boundary (twist angle ~ 8.6°). The π energy loss peak MoTe2 locates at 6.4 eV (2H) and 5.7 eV (1T'). The energy position of π+σ peak is about 19 eV. (b) the dielectric function of 2H-MoTe2. (c) the calculated loss function based on dielectric function in (a).

The longitudinal wavelike oscillations of MoTe2 weakly bound electrons generate the plasmon and cause the energy loss.

The energy loss function $L$ is calculated on the grounds of the function:

$$L = -I_m\left[\frac{1}{\varepsilon(\omega)}\right]$$

where $\varepsilon(q, \omega)$ is the dielectric function produced by first principles calculations. The energy loss function of 2H-MoTe2 incorporates π (6.4 eV) and π+σ (19.1 eV) peaks, which are consistent with experimental observation.

Two dominated peaks π and π+σ can be observed in the energy loss range of 5-35 eV. The π plasmon mode in MoTe2 results from the π-π* transitions, while π+σ plasmon mode arises from the π-σ* and σ-σ* excitations. [3]

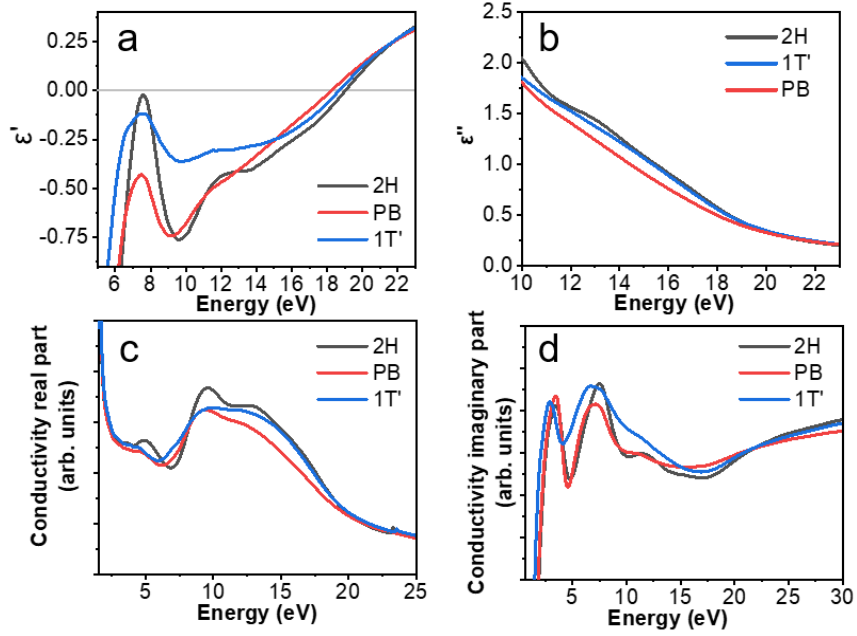

**FIG. S5. The K-K analysis of plasmon mode of 1T'/2H-MoTe$_2$ phase boundary.** (a) The real part and (b) the imaginary part of the dielectric function of 2H-MoTe$_2$, 1T'-MoTe$_2$, and at the phase boundary are indicated by black, blue and red, respectively. (c) The real part and (d) the imaginary part of conductivity deduced from the dielectric function.

The increased real part and the decreased imaginary part of the dielectric function at phase boundary (in the range of 15-21 eV) contributes to the energy red shift and intensity decrease of π+σ peak. The conductivity at the phase boundary slightly decreased compared to the intrinsic MoTe$_2$.

The red shift of π+σ plasmon mode at the phase boundary is attributed to the reduction of effective electron density at the phase boundary. The free electron density valley at the phase boundary might generate a barrier for the carrier injection from the 1T'-MoTe$_2$ to 2H-MoTe$_2$, which is detrimental for contact.

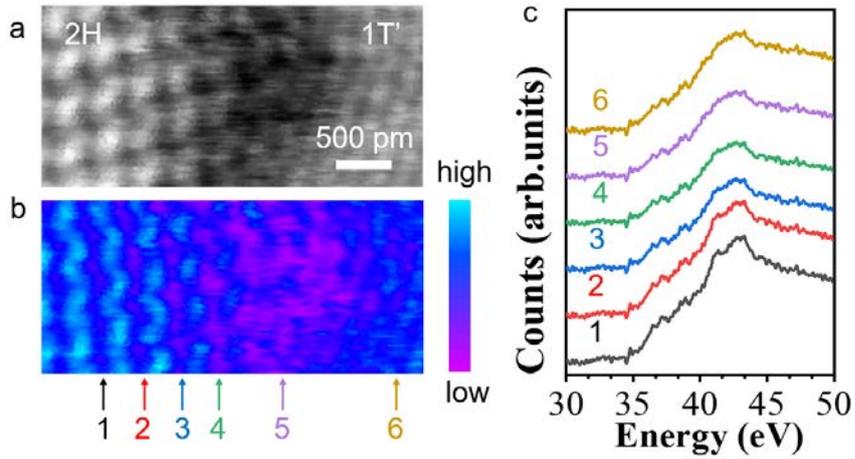

**FIG. S6. Te-N ELNES at GB of 8.6° tilt angle.** (a) A HAADF image showing the spectra-collected region. (b) Spectral image of the Te-N edge. The spectral image is extracted by applying an integral window of 30-50 eV. (c) Te-N edges of six locations are indicated by black, red, blue, green, purple and yellow arrows in (b), respectively.

In this case, the Te-N edge of 2H-MoTe$_2$ also changed within 2 unit cells at the 8.6° tilted boundary, which shows the similar behaviors as the ~0° tilted boundary.

It is worth mentioning that the typical probe size is ~1.07 Å which can be measured based on the resolution of atomically resolved image. As a result, the sampling region of atomically resolved STEM-EELS can be very local.

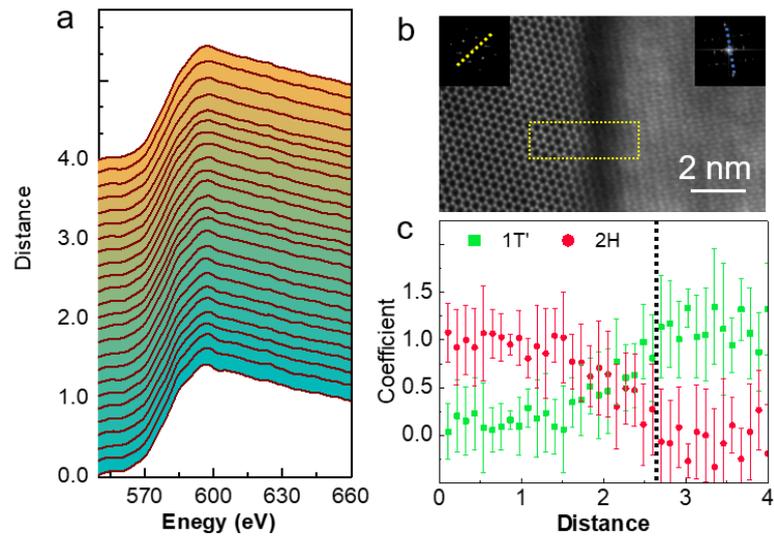

**FIG. S7. Te-M ELNES at GB of 53.5° tilt angle.** (a) Spectra series across the 1T'/2H phase boundary. (b) A HAADF image showing the 53.5 tilted phase boundary. (c) Fitting coefficient of Te-M at 1T'/2H-MoTe$_2$ boundary plotted as a function of distance. Red sphere: 2H; Green square: 1T'.

## Experimental Methods

**Synthesis of the MoTe$_2$ Films.** The MoTe$_2$ films were synthesized by tellurizing the Mo film at atmospheric pressure using a horizontal hot-wall tube furnace equipped with mass flow controllers and a vacuum pump. Mo films were deposited on Si/SiO$_2$ substrates through magnetron sputtering. The substrates were placed face-down on an alumina boat containing Te powder placed at the center of the heating zone in a one-inch quartz tube. After evacuating the quartz tube to less than 1 mTorr, we flowed Ar gas at 500 standard cubic centimeters per minute (sccm) until the pressure reached atmospheric pressure. At atmospheric pressure, Ar and H$_2$ flowed at rates of 4 and 5 sccm, respectively. The furnace was ramped to 650 °C in 15 min and was kept at the temperatures for 30 min to synthesize the coplanar contact structure. After the reactions, we let the furnace cool to room temperature naturally.

**Transfer of the 1T′/2H hetero-phase MoTe$_2$ thin film.** First, drop a few mL of isopropanol (IPA) onto the hetero-phase MoTe$_2$ thin film on the Si/SiO$_2$ substrate. A mesh copper grid was carefully put onto the IPA solution. After the IPA evaporates naturally, the copper grid is glued to the hetero-phase MoTe$_2$ thin film. Then, a small amount of the dilute HF solution (1.5%) was dropped onto the edge of the copper grid. A few seconds later, the copper grid along with the MoTe$_2$ film floats in solution. Finally, the sample was thoroughly rinsed with DI water.

**STEM and EELS characterization.** The STEM-HAADF images were acquired using a Nion HERMES 200 microscope with both monochromator and the aberration corrector operating at 60 kV. The convergence semi-angle was 35 mrad and the collection semi-angle was in the range of 80-210 mrad. In order to achieve high spatial resolution, the EEL spectrum images were recorded at STEM mode with a 1 mm spectrometer entrance aperture. The STEM-EELS were recorded at 60 kV with a collection semi-angle of 24.9 mrad. The EELS background was fitted and subtracted using power law $I(\Delta E) = A_0 \cdot \Delta E^{-r}$. The interfacial VEELS were fitted by multiple linear least-squares (MLLS) method using DigitalMicrograph (Gatan) software. The intrinsic VEELS of 1T'-MoTe$_2$ and 2H-MoTe$_2$ and were used as reference spectra respectively. The spectra were normalized using the zero-loss peak of EELS. The low loss plasmon EELS spectrum for K-K analysis was Fourier-log deconvolved in order to extract the single scattering distribution [4].

**DFT calculation.** The geometry optimizations of both 1T' and 2H-MoTe$_2$ have been fully calculated before the electronic structure calculations. The plane-wave basis set with a cut-off energy of 400 eV and the projector augmented wave (PAW) pseudopotential are used in the Vienna *ab initio* simulation package (VASP). Throughout the calculations, the generalized gradient approximation (GGA) combined with the Perdew-Burke-Ernzerhof (PBE) form are adopted. The maximum residual force of per atom is less than 0.001 eV/Å to obtain a reliable optimized structure, and the convergence standard of energy on each atom is within $1 \times 10^{-6}$ eV. We use the Monkhorst–Pack method to sample the *k*-point mesh with a separation of 0.04 Å$^{-1}$ and 0.02 Å$^{-1}$ in the irreducible Brillouin zone (IBZ) for the geometry optimizations and electronic structure calculations, respectively.